\documentstyle[epsf,tighten,preprint,prd,aps]{revtex}

\begin{document}
\epsfverbosetrue
\draft


\title{
  Self-Similar Collapse of Scalar Field in Higher Dimensions
}
\author{Andrei V. Frolov%
  \thanks{Email: \texttt{andrei@phys.ualberta.ca}}}
\address{
  Physics Department, University of Alberta\\
  Edmonton, Alberta, Canada, T6G 2J1}
\date{June 28, 1998}
\maketitle

\begin{abstract}
  This paper constructs continuously self-similar solution of a
  spherically symmetric gravitational collapse of a scalar field in $n$
  dimensions. The qualitative behavior of these solutions is explained,
  and closed-form answers are provided where possible. Equivalence of
  scalar field couplings is used to show a way to generalize minimally
  coupled scalar field solutions to the model with general coupling.
\end{abstract}

\pacs{PACS numbers: 04.70.Bw, 05.70.Jk}
\narrowtext


\section{Introduction} \label{sec:intro}

Choptuik's discovery of critical phenomena in gravitational collapse of
a scalar field \cite{Choptuik:93} sparkled a surge of interest in
gravitational collapse just at the threshold of black hole formation.
Discovery of critical behavior in several other matter models quickly
followed \cite{Abrahams&Evans:93,Evans&Coleman:94,Koike&Hara&Adachi:95,Maison:96,Hirschmann&Eardley:95a,Hirschmann&Eardley:95b}.
While perhaps the presence of critical behavior in gravitational
collapse is not in itself surprising, some of its features are, in
particular the conclusion that black holes of arbitrary small mass can
be formed in the process. Moreover, the critical solution often
displays additional peculiar symmetry -- the so-called self-similarity,
and serves as an intermediate attractor for near-critical solutions.

Study of critical phenomena also throws new light on the cosmic
censorship conjecture. The formation of strong curvature singularity in
critical collapse from regular initial data offers a new counterexample
to the cosmic censorship conjecture.

Much work has been done, and general features of critical behavior are
now understood. However, there is a distinct and uncomfortable lack of
analytical solutions. Due to the obvious difficulties in obtaining
solutions of Einstein equations in closed form, most of the work seems
to be done numerically. One of the few known closed form solutions
related to critical phenomena is Roberts solution, originally
constructed as a counterexample to the cosmic censorship conjecture
\cite{Roberts:89}, and later rediscovered in the context of critical
gravitational collapse \cite{Brady:94,Oshiro&Nakamura&Tomimatsu:94}. It
is a continuously self-similar solution of a spherically symmetric
gravitational collapse of a minimally coupled massless scalar field in
four dimensional spacetime. For review of the role self-similarity
plays in general relativity see \cite{Carr&Coley:98}.

This paper searches for continuously self-similar, spherically
symmetric scalar field solutions in $n$ dimensions. They might be
relevant in context of the superstring theory, which is often said to
be the next ``theory of everything'', as well as for understanding how
critical behavior depends on the dimensionality of the spacetime.
Roberts solution would be a particular case of the solutions discussed
here. These solutions provide reasonably simple toy models of critical
collapse, although they are not attractors \cite{Frolov:97}. Some
qualitative properties of the self-similar critical collapse of scalar
field in higher dimensions have been discussed in
\cite{Soda&Hirata:96}. Here we aim at finding explicit closed-form
solutions.

Second part of this paper deals with extension of minimally coupled
scalar field solutions to a wider class of couplings. It is shown how
different couplings of scalar field are equivalent, and several
particular models are examined in details. Procedure discussed here can
be applied to any solution of Einstein-scalar field equations.

\section{Reduced Action and Field Equations} \label{sec:action}

Evolution of the minimally coupled scalar field in $n$ dimensions is
described by action
\begin{equation} \label{eq:action}
  S = \frac{1}{16\pi} \int \sqrt{-g}\, d^n\!x \left[R - 2 (\nabla\phi)^2\right]
\end{equation}
plus surface terms. Field equations are obtained by varying this action
with respect to field variables $g_{\mu\nu}$ and $\phi$. However, if
one is only interested in spherically symmetric solutions (as we are),
it is much simpler to work with reduced action and field equations,
where this symmetry of the spacetime is factored out.

Spherically symmetric spacetime is described by the metric
\begin{equation} \label{eq:action:metric}
  ds^2 = d\gamma^2 + r^2 d\omega^2,
\end{equation}
where $d\gamma^2$ is the metric on a two-manifold, and $d\omega^2$ is
the metric of a $(n-2)$-dimensional sphere. Essentially, the spherical
symmetry reduces the number of dimensions to two, with spacetime fully
described by the two-metric $d\gamma^2$ and two-scalar $r$. It can be
shown that the reduced action describing field dynamics in spherical
symmetry is
\begin{equation} \label{eq:action:reduced}
  S_{\text{sph}} \propto
    \int r^{n-2} \sqrt{-\gamma}\, d^2\!x \left[
      R[\gamma] + (n-2)(n-3) r^{-2} \left( (\nabla r)^2 + 1 \right)
      - 2 (\nabla\phi)^2
    \right],
\end{equation}
where curvature and differential operators are calculated using
two-dimensional metric $\gamma_{AB}$; capital Latin indices run through
$\{1, 2\}$, and stroke $|$ denotes covariant derivative with respect to
two-metric $\gamma_{AB}$. Varying reduced action with respect to field
variables $\gamma_{AB}$, $r$, and $\phi$, we obtain Einstein-scalar
field equations in spherical symmetry. After some algebraic
manipulation they can be written as
\begin{mathletters} \label{eq:einstein}
\begin{equation} \label{eq:einstein:R}
  R_{AB} - (n-2) r^{-1} r_{|AB} = 2 \phi_{,A} \phi_{,B},
\end{equation}
\begin{equation} \label{eq:einstein:r}
  (n-3) \left[ (\nabla r)^2 - 1 \right] + r \Box r = 0,
\end{equation}
\begin{equation} \label{eq:einstein:phi}
  \Box\phi + (n-2) r^{-1} \gamma^{AB} \phi_{,A} r_{,B} = 0.
\end{equation}
\end{mathletters}
As usual with scalar field, $\Box\phi$ equation is redundant.

\section{$n$-dimensional Generalization of Roberts Solution} \label{sec:soln}

We are interested in generalization of Roberts solution to $n$
dimensions. To find it, we write the metric in double null coordinates
\begin{equation} \label{eq:soln:metric}
  d\gamma^2 = -2 e^{-2\sigma(z)} du\,dv, \hspace{1cm}
  r = - u \rho(z), \hspace{1cm}
  \phi = \phi(z).
\end{equation}
The dependence of metric coefficients and $\phi$ on $z= -v/u$ only
reflects the fact that we are looking for continuously self-similar
solution, with $z$ being scale-invariant variable. We turn on the
influx of the scalar field at the advanced time $v=0$, so that the
spacetime is Minkowskian to the past of this surface, and the initial
conditions are specified by continuity. Signs are chosen so that $z>0$,
$\rho>0$ in the sector of interest ($u<0$, $v>0$). With this choice of
metric, Einstein-scalar equations (\ref{eq:einstein}) become
\begin{mathletters} \label{eq:soln:einstein}
\begin{equation} \label{eq:soln:einstein:R11}
  (n-2) \left[ \rho'' z + 2 \sigma' \rho - 2 \sigma' \rho' z \right]
    = - 2 \rho z {\phi'}^2,
\end{equation}
\begin{equation} \label{eq:soln:einstein:R12}
  2 \rho (\sigma'' z + \sigma') + (n-2) \rho'' z = - 2 \rho z {\phi'}^2,
\end{equation}
\begin{equation} \label{eq:soln:einstein:R22}
  (n-2) \left[ \rho'' - 2 \sigma' \rho' \right] = - 2 \rho {\phi'}^2,
\end{equation}
\begin{equation} \label{eq:soln:einstein:r}
  (n-3) \left[ {\rho'}^2 z - \rho' \rho + \frac{1}{2}\, e^{2\sigma} \right]
    + \rho'' \rho z = 0,
\end{equation}
\begin{equation} \label{eq:soln:einstein:phi}
  \phi'' \rho z + (n-2) \phi' \rho' z - \frac{1}{2}\, (n-4) \phi' \rho = 0.
\end{equation}
\end{mathletters}
Prime denotes the derivative with respect to $z$. Combining equations
(\ref{eq:soln:einstein:R11}) and (\ref{eq:soln:einstein:R22}), we obtain
that $\sigma = \text{const}$. By appropriate rescaling of coordinates,
we can put $\sigma = 0$. Then
\begin{mathletters} \label{eq:soln:eqn}
\begin{equation} \label{eq:soln:eqn:R}
  (n-2) \rho'' = - 2 \rho {\phi'}^2,
\end{equation}
\begin{equation} \label{eq:soln:eqn:r}
  (n-3) \left[ {\rho'}^2 z - \rho' \rho + \frac{1}{2} \right]
    + \rho'' \rho z = 0,
\end{equation}
\begin{equation} \label{eq:soln:eqn:phi}
  \frac{\phi''}{\phi'} + (n-2) \frac{\rho'}{\rho}
    - \frac{1}{2}\, (n-4) z^{-1} = 0.
\end{equation}
\end{mathletters}
For further derivation we will assume that $n>3$, as the case $n=3$ is
trivial. Equation (\ref{eq:soln:eqn:phi}) can be immediately integrated
\begin{equation} \label{eq:soln:int:phi}
  \phi' \rho^{n-2} z^{-(n-4)/2} = c_0.
\end{equation}
Substituting this result back into equation (\ref{eq:soln:eqn:R}), we
get the equation for $\rho$ only
\begin{equation} \label{eq:soln:int:rho}
  \rho'' \rho^{2n-5} = - \frac{2 c_0^2}{n-2}\, z^{n-4}.
\end{equation}
It is easy to show that equation (\ref{eq:soln:int:rho}) is equivalent
to equation (\ref{eq:soln:eqn:r}). No surprises here, since the system
(\ref{eq:einstein}) was redundant. Combining both equations we get first
integral of motion
\begin{equation} \label{eq:soln:int}
  \left[ {\rho'}^2 z - \rho' \rho + \frac{1}{2} \right]
    \left( \frac{\rho^2}{z} \right)^{n-3} =
    \frac{2 c_0^2}{(n-2)(n-3)},
\end{equation}
which contains only first derivatives of $\rho$, and for this reason is
simpler to solve than either one of equations (\ref{eq:soln:eqn:r}) and
(\ref{eq:soln:int:rho}). Equation (\ref{eq:soln:int}) is generalized
homogeneous one, and can be solved by substitution
\begin{equation} \label{eq:soln:subst}
  x = \frac{1}{2} \ln z, \hspace{1cm}
  \rho = \sqrt{z} y(x), \hspace{1cm}
  \rho' = \frac{1}{2}\, z^{-1/2} (\dot{y} + y),
\end{equation}
where dot denotes the derivative with respect to new variable $x$. With
this substitution, equations (\ref{eq:soln:int}) and
(\ref{eq:soln:int:phi}) become
\begin{equation} \label{eq:soln:y}
  \dot{y}^2 = y^2 - 2 + c_1 y^{-2(n-3)},
\end{equation}
\begin{equation} \label{eq:soln:phi}
  \dot{\phi} = 2 c_0 y^{-(n-2)},
\end{equation}
where we redefined constant
\begin{equation} \label{eq:soln:c1}
  c_1 = \frac{8 c_0^2}{(n-2)(n-3)} > 0.
\end{equation}
The above equation (\ref{eq:soln:y}) for $y$ formally describes motion
of a particle with zero energy in potential
\begin{equation} \label{eq:soln:potential}
  V(y) = 2 - y^2 - c_1 y^{-2(n-3)},
\end{equation}
so we can tell qualitative behavior of $y$ without actually solving
equation (\ref{eq:soln:y}).

\begin{figure}
  \centerline{
    \epsfxsize=\columnwidth \multiply\epsfxsize 4 \divide\epsfxsize 5
    \epsfbox{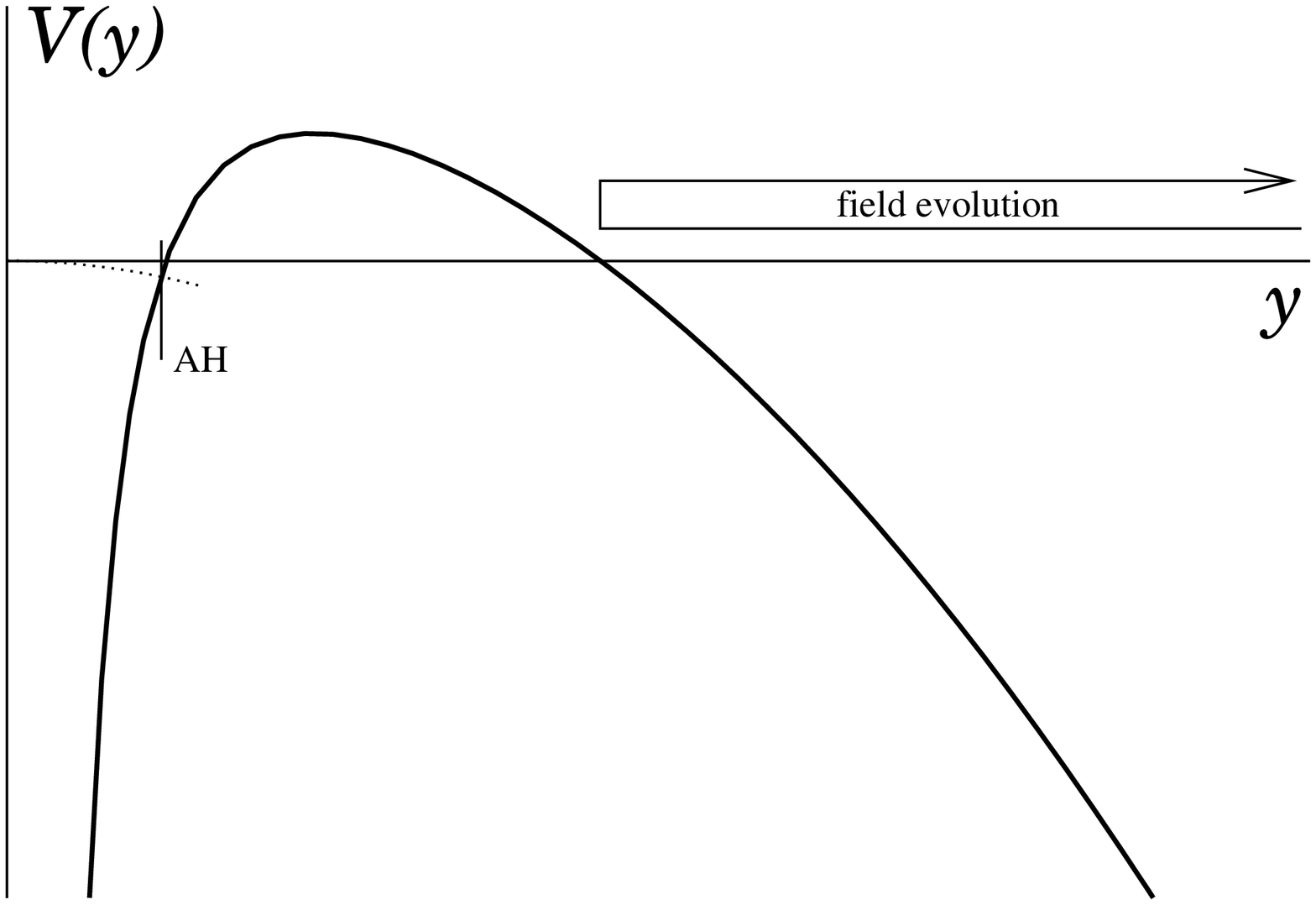}
  }
  \caption{Subcritical field evolution.}
  \label{fig:subcritical}
\end{figure}

\begin{figure}
  \centerline{
    \epsfxsize=\columnwidth \multiply\epsfxsize 4 \divide\epsfxsize 5
    \epsfbox{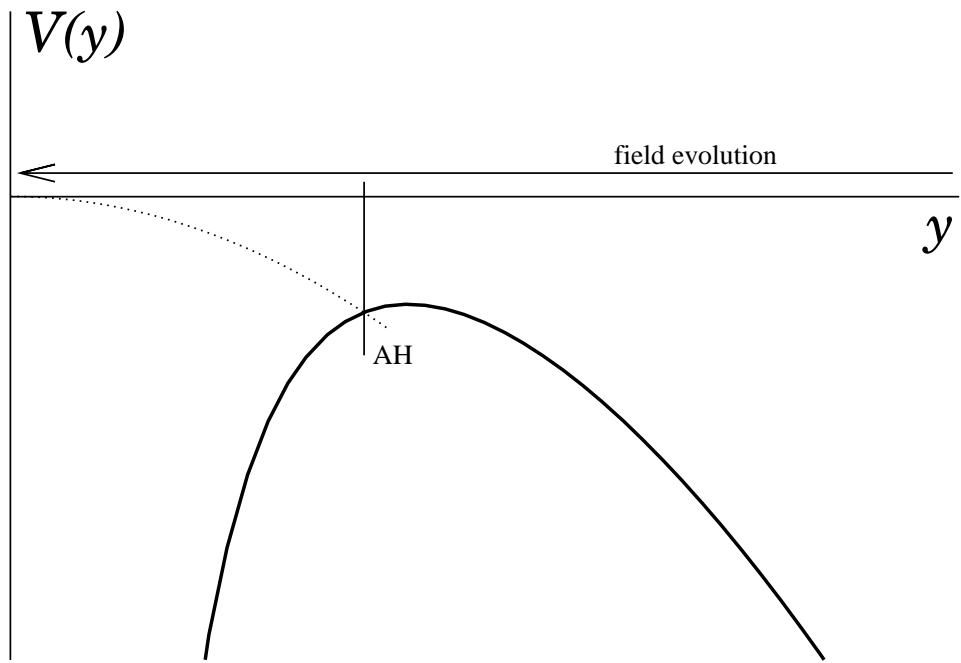}
  }
  \caption{Supercritical field evolution.}
  \label{fig:supercritical}
\end{figure}

\begin{figure}
  \centerline{
    \epsfxsize=\columnwidth \multiply\epsfxsize 4 \divide\epsfxsize 5
    \epsfbox{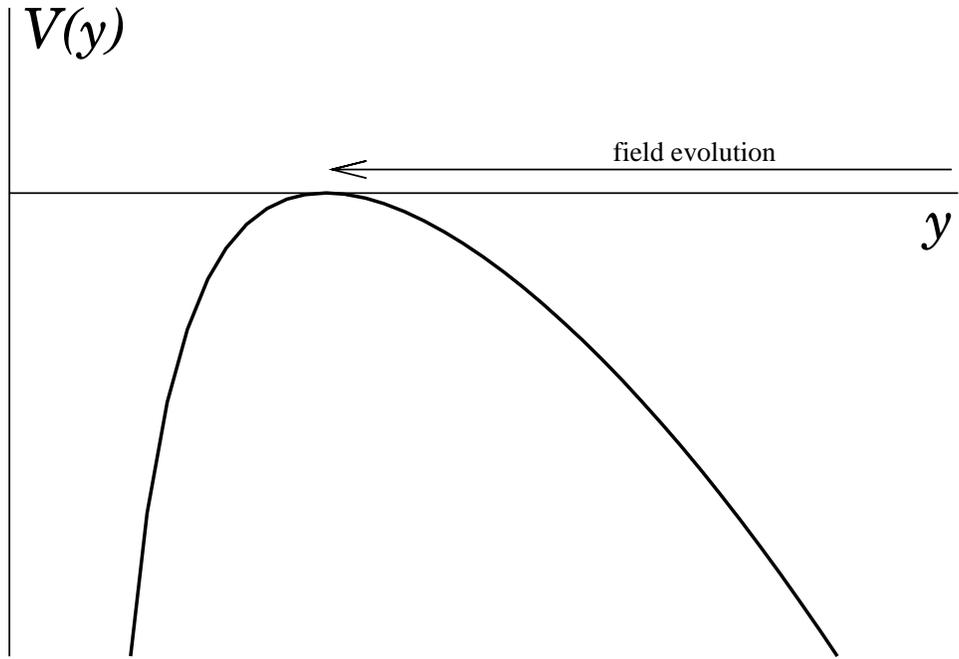}
  }
  \caption{Critical field evolution.}
  \label{fig:critical}
\end{figure}

Initial conditions are specified by continuous matching of the solution
to Minkowskian spacetime on surface $v=0$. Since on that surface
$r \neq 0$, the value of $y=r/\sqrt{-uv}$ starts from infinity at
$x=-\infty$, and rolls towards zero. What happens next depends on the
shape of the potential. If there is region with $V(y)>0$, as in
Fig.~\ref{fig:subcritical}, $y$ will reach a turning point and will go
back to infinity as $x=\infty$. If $V(y)<0$ everywhere, as in
Fig.~\ref{fig:supercritical}, there is nothing to stop $y$ from
reaching zero, at which point singularity is formed. Finally, if $V(y)$
has second-order zero, as in Fig.~\ref{fig:critical}, $y$ will take
forever reaching it.

Of course, variables separate, and equation (\ref{eq:soln:y}) can be
integrated
\begin{equation} \label{eq:soln}
  x = \pm \int \frac{dy}{\sqrt{y^2 - 2 + c_1 y^{-2(n-3)}}} + c_2.
\end{equation}
Plus or minus sign in front of the integral depends on the sign of the
derivative of $y$. Initial conditions imply that initially $y$ comes
from infinity towards zero, i.e.\ its derivative is negative, and so we
must pick the branch of the solution which started out with a minus
sign. Constant $c_2$ corresponds to a coordinate freedom in choice of
origin of $x$, while constant $c_1$ is a real parameter of the
solution.

Unfortunately, the integral cannot be evaluated in a closed form for
arbitrary $n$. But if the integral is evaluated, and we can invert it to
get $y$ as a function of $x$, the solution for $r$ is obtained by using
definitions (\ref{eq:soln:subst}) and (\ref{eq:soln:metric}). The
solution for $\phi$ is obtained by integrating relations
(\ref{eq:soln:phi}) or (\ref{eq:soln:int:phi}).

\section{Critical Behavior} \label{sec:critical}

One-parameter family of self-similar scalar field solutions in $n$
dimensions we constructed above exhibits critical behavior as the
parameter $c_1$ is tuned, much as Roberts family does in four
dimensions. In this section we investigate black hole formation in the
collapse.

In spherical symmetry, existence and position of the apparent horizon
are given by vanishing of $(\nabla r)^2=0$, which translates to
$\rho'=0$, or $\dot{y}+y=0$ in our notation. Therefore, at the apparent
horizon we have
\begin{equation}
  \dot{y}^2 - y^2 = c_1 y^{-2(n-3)} - 2 = 0,
\end{equation}
and the black hole is formed if value of $y$ reaches
\begin{equation} \label{eq:crit:y:ah}
  y_{\text{AH}}^2 = \left(\frac{c_1}{2}\right)^{\frac{1}{n-3}}.
\end{equation}
As we have discussed above, depending on the value of $c_1$, values of
field $y$ either reach turning point and return to infinity, or go all
the way to zero. The critical solution separates the two cases, and is
characterized by potential $V(y)$ having second order zero, i.e.\ 
$V(y_*)=V'(y_*)=0$ at some point $y_*$. Differentiating expression
(\ref{eq:soln:potential}) for potential $V(y)$, we see that it has
second order zero at
\begin{equation} \label{eq:crit:y:zero}
  y_*^2 = 2\, \frac{n-3}{n-2},
\end{equation}
if and only if the value of the constant $c_1$ is
\begin{equation} \label{eq:crit:c1}
  c_1^* = \frac{1}{n-3} \left[2\, \frac{n-3}{n-2}\right]^{n-2}.
\end{equation}
If the value of parameter $c_1$ is less than critical, $c_1<c_1^*$, the
value of $y$ turns around at the turning point, and never reaches point
of the apparent horizon formation, which is located in the forbidden
zone, as illustrated in Fig.~\ref{fig:subcritical}. This case is
subcritical evolution of the field. If $c_1>c_1^*$, the value of $y$
reaches point where apparent horizon is formed, and proceeds to go to
zero, at which place there is a singularity inside the black hole. This
supercritical evolution is illustrated in Fig.~\ref{fig:supercritical}.

The mass of the black hole formed in the supercritical collapse is
\begin{equation} \label{eq:crit:mass}
  M = \frac{1}{2}\, r_{\text{AH}}
    = - \frac{1}{2}\, u \sqrt{z_{\text{AH}}}\, y_{\text{AH}}.
\end{equation}
It grows infinitely if we wait long enough, and will absorb all the
field influx coming from past infinity. This happens because the
solution is self-similar, and creates a problem for discussing mass
scaling in the near-critical collapse. Cut and glue schemes
\cite{Wang&Oliveira:96} avoiding infinite black hole mass are temporary
means to lift this problem. However, the real answer to determining if
the critical solution is intermediate attractor and calculating
mass-scaling exponent is to make a perturbative analysis of critical
solution. Similarity to Roberts solution suggests that the results for
four dimensional spacetime \cite{Frolov:97} can be applied to higher
dimensions as well.

\section{Particular Cases} \label{sec:particular}

In this section we consider several particular cases for which the
general solution (\ref{eq:soln}) is simplified. Particularly important
is $n=4$ case, which is the already familiar Roberts solution.

\subsection{$n=3$} \label{sec:roberts:3}

As we have already mentioned, for $n=3$ the only self-similar scalar
field solution of the form (\ref{eq:soln:metric}) is trivial. To see it,
note that equation (\ref{eq:soln:eqn:r}) implies that $\rho''=0$ if
$n=3$, and so $\rho = \alpha z + \beta$ and $r = \alpha v - \beta u$.
From equation (\ref{eq:soln:eqn:R}) it then follows that $\phi =
\text{const}$. The spacetime is flat.

\subsection{$n=4$} \label{sec:roberts:4}

Integration (\ref{eq:soln}) can be carried out explicitly
\begin{eqnarray}
  x &=& - \int \frac{dy}{\sqrt{y^2 - 2 + c_1 y^{-2}}} + c_2 \nonumber\\
    &=& - \frac{1}{2} \ln \left|y^2-1 + \sqrt{y^4-2y^2+c_1}\right| + c_2,
\end{eqnarray}
and the result inverted
\begin{equation}
  y^2 = \frac{1}{2} e^{-2(x-c_2)} + 1 + \frac{1-c_1}{2}\, e^{2(x-c_2)},
\end{equation}
to give the solution in the closed form
\begin{equation}
  \rho = \sqrt{\frac{e^{2c_2}}{2} + z + \frac{1-c_1}{2 e^{2c_2}}\, z^2}.
\end{equation}
By appropriately rescaling coordinates, we can put $e^{2c_2}=2$.
After redefining parameter of the solution $p=(c_1-1)/4$, the solution
takes on the following simple form
\begin{equation}
  \rho = \sqrt{1 + z - p z^2}, \hspace{1cm}
  r = \sqrt{u^2 - uv - p v^2}.
\end{equation}
Scalar field $\phi$ is reconstructed from equation
(\ref{eq:soln:int:phi})
\begin{equation}
  \phi' = c_0 \rho^{-2}
        = \frac{1}{2}\, \frac{\sqrt{1+4p}}{1 + z - p z^2},
\end{equation}
to give
\begin{eqnarray}
  \phi &=& \text{arctanh}\, \frac{2pz-1}{\sqrt{1+4p}}
             + \text{const} \nonumber\\
       &=& \frac{1}{2}\, \ln \left[ - \frac{2pz-1 + \sqrt{1+4p}}
               {2pz-1 - \sqrt{1+4p}} \right] + \text{const}.
\end{eqnarray}
The critical parameter value is $p^*=0$, and for $p>0$ the black hole
is formed. The critical solution is
\begin{equation}
  r = \sqrt{u^2 - uv}, \hspace{1cm}
  \phi = \frac{1}{2} \ln \left[1 - \frac{v}{u}\right].
\end{equation}

\subsection{$n=5, 6$} \label{sec:roberts:5,6}

Integral (\ref{eq:soln}) can be written in terms of elliptic functions,
which becomes apparent with the change of variable $\bar{y}=y^{-2}$
\begin{equation}
  x = \frac{1}{2} \int \frac{d\bar{y}}
        {\bar{y} \sqrt{1-2\bar{y}+c_1 \bar{y}^{n-2}}},
\end{equation}
and the solution $y(x)$ is given implicitly. However, integrals
corresponding to critical solutions simplify, and can be taken in terms
of elementary functions for $n=5,6$. It happens because potential
factors since it has second order zero, therefore reducing the power of
$y$ in the radical by two. The results of integration for critical
solutions are
\begin{eqnarray}
  n=5: \ \ 
  x &=& - \int \frac{y^2 dy}{\left(y^2 - \frac{4}{3}\right) 
                             \sqrt{y^2 + \frac{2}{3}}} \nonumber\\
    &=& -\, \text{arcsinh} \left(\sqrt{\frac{3}{2}}\, y\right)
        + \frac{1}{\sqrt{6}}\ \text{arctanh} 
	     \left(\frac{\sqrt{3}\, y + 1}{\sqrt{\frac{9}{2}\,y^2+3}}\right)
        + \frac{1}{\sqrt{6}}\ \text{arctanh} 
	     \left(\frac{\sqrt{3}\, y - 1}{\sqrt{\frac{9}{2}\,y^2+3}}\right) \\
  n=6: \ \ 
  x &=& - \int \frac{y^3 dy}{\left(y^2 - \frac{3}{2}\right)
                             \sqrt{y^4 + y^2 + \frac{3}{4}}} \nonumber\\
    &=& -\, \frac{1}{2}\ \text{arcsinh} \left(\sqrt{2} (y^2+\frac{1}{2})\right)
        + \frac{1}{2\sqrt{2}}\ \text{arctanh}
	     \left(\frac{1}{\sqrt{2}}\, \frac{\frac{4}{3}\, y^2 + 1}
	            {\sqrt{y^4+y^2+\frac{3}{4}}}\right).
\end{eqnarray}
The dependence $y(x)$ is still given implicitly. The critical value of
the parameter $c_1^*$ is $32/27$ for $n=5$, and $27/16$ for $n=6$.

\subsection{Higher dimensions} \label{sec:roberts:>6}

For higher dimensions, integral (\ref{eq:soln}) can not be taken in
terms of elementary functions, so one has to be content with the
solution in the integral-implicit form, or do numerical calculations.

\section{General Scalar Field Coupling} \label{sec:coupling}

In this section we discuss in detail how solutions of minimally coupled
scalar field model can be generalized to much wider class of couplings.
The fact that essentially all couplings of a free scalar field to its
kinetic term and scalar curvature are equivalent has been used in the
past \cite{Bekenstein:74,Page:91} to study scalar field models with
non-minimal coupling. In particular, this idea has been applied to
extend four-dimensional Roberts solution to conformal coupling
\cite{Oliveira&Cheb-Terrab:96} and Brans-Dicke
theory \cite{Oliveira:96}.

\subsection{Equivalence of Couplings} \label{sec:coupling:equiv}

\newcommand{\hBox}{\hat{\raisebox{0pt}[1.4\height]{$\Box$}}}

Suppose that the action describing evolution of the scalar field in
$n$-dimensional spacetime is
\begin{equation} \label{eq:coupling:general}
  S = \frac{1}{16\pi} \int \sqrt{-g}\, d^n\!x \left[
        F(\phi) R - G(\phi) (\nabla\phi)^2
      \right]
\end{equation}
plus surface terms, where the couplings $F$ and $G$ are smooth
functions of field $\phi$. Also suppose that the signs of couplings $F$
and $G$ correspond to the case of gravitational attraction. We will
demonstrate that this action reduces to the minimally coupled one by
redefinition of the fields $g_{\mu\nu}$ and $\phi$. First, let's
introduce new metric $\hat{g}_{\mu\nu}$ that is related to old one by
conformal transformation
\begin{equation} \label{eq:coupling:metric}
  \hat{g}_{\mu\nu} = \Omega^2 g_{\mu\nu}, \hspace{1cm}
  \hat{g}^{\mu\nu} = \Omega^{-2} g^{\mu\nu}, \hspace{1cm}
  \sqrt{-\hat{g}} = \Omega^n \sqrt{-g},
\end{equation}
and denote quantities and operators calculated using $\hat{g}_{\mu\nu}$
by hat. Scalar curvatures calculated using old and new metrics are
related
\begin{equation} \label{eq:coupling:R}
  R = \Omega^2 \hat{R}
      + 2(n-1) \Omega \hBox\Omega
      - n(n-1) (\hat{\nabla}\Omega)^2,
\end{equation}
as are field gradients
\begin{equation} \label{eq:coupling:gradient}
  (\nabla\phi)^2 = \Omega^2 (\hat{\nabla}\phi)^2.
\end{equation}
Writing action (\ref{eq:coupling:general}) in terms of metric
$\hat{g}_{\mu\nu}$, we obtain
\begin{equation} \label{eq:coupling:S1}
  S = \frac{1}{16\pi} \int \Omega^{-n} \sqrt{-\hat{g}}\, d^n\!x \left[
        F \{\Omega^2 \hat{R} + 2(n-1) \Omega \hBox\Omega
	    - n(n-1) (\hat{\nabla}\Omega)^2\}
	- G \Omega^2 (\hat{\nabla}\phi)^2
      \right].
\end{equation}
By choosing conformal factor to be
\begin{equation} \label{eq:coupling:omega}
  \Omega^{n-2} = F,
\end{equation}
the factor in front of the curvature $\hat{R}$ can be set to one.
Substitution of definition of $\Omega$ into the above action, and
integration by parts of $\hBox$ operator yields
\begin{equation} \label{eq:coupling:S2}
  S = \frac{1}{16\pi} \int \sqrt{-\hat{g}}\, d^n\!x \left[
        \hat{R} - \left( \frac{G}{F} + \frac{n-1}{n-2}\,
	  \frac{{F'}^2}{F^2}\right) (\hat{\nabla}\phi)^2
      \right].
\end{equation}
The kinetic term in action (\ref{eq:coupling:S2}) can be brought into
minimal form by introduction of a new scalar field $\hat{\phi}$,
related to the old one by
\begin{equation} \label{eq:coupling:phi}
  2 (\hat{\nabla}\hat{\phi})^2 =
    \left(\frac{G}{F} + \frac{n-1}{n-2}\,
    \frac{{F'}^2}{F^2}\right) (\hat{\nabla}\phi)^2.
\end{equation}
Thus, we have shown that with field redefinitions
\begin{equation} \label{eq:coupling:redefinition}
  \hat{\phi} = \frac{1}{\sqrt{2}} \int
    \left(\frac{G}{F} + \frac{n-1}{n-2}\,
    \frac{{F'}^2}{F^2}\right)^{\frac{1}{2}} d\phi,
  \hspace{1cm}
  \hat{g}_{\mu\nu} = F^{2/(n-2)} g_{\mu\nu},
\end{equation}
the generally coupled scalar field action (\ref{eq:coupling:general})
becomes minimally coupled
\begin{equation} \label{eq:coupling:minimal}
  S = \frac{1}{16\pi} \int \sqrt{-\hat{g}}\, d^n\!x \left[
        \hat{R} - 2(\hat{\nabla}\hat{\phi})^2
      \right].
\end{equation}
This equivalence allows one to construct solutions of the model with
general coupling (\ref{eq:coupling:general}) from the solutions of
minimally coupled scalar field by means of inverse of relation
(\ref{eq:coupling:redefinition}), provided said inverse is
well-defined. However, there may be some restrictions on the range of
$\phi$ so that field redefinitions give real $\hat{\phi}$ and
positive-definite metric $\hat{g}_{\mu\nu}$. This means that not all
the branches of the solution in general coupling may be covered by
translating minimally coupled solution. Technically speaking, the
correspondence between solutions of minimally coupled theory and
generally coupled theory is one-to-one where defined, but not onto.

However, one has to be careful making claims about global structure and
critical behavior of the generalized solutions based solely on the
properties of the minimally-coupled solution. The scalar field solutions
encountered in critical collapse often lead to singular conformal
transformations, which could, in principle, change the structure of
spacetime.

\subsection{Examples} \label{sec:coupling:examples}

To illustrate the above discussion, we consider two often used scalar
field couplings as examples. They are non-minimal coupling and dilaton
gravity.

\subsubsection{Conformal coupling} \label{sec:coupling:conformal}
Non-minimally coupled scalar field in $n$ dimensions is described by
action
\begin{equation} \label{eq:conformal:action}
  S = \frac{1}{16\pi} \int \sqrt{-g}\, d^n\!x \left[
        (1-2\xi\phi^2) R - 2 (\nabla\phi)^2
      \right],
\end{equation}
where $\xi$ is the coupling parameter. Coupling factors are
$F=1-2\xi\phi^2$, $G=2$ and so the field redefinition
(\ref{eq:coupling:redefinition}) looks like
\begin{eqnarray}
  \hat{\phi} &=& \int \frac{\sqrt{1 - 2\xi\phi^2 + 2 \xi_c^{-1} \xi^2 \phi^2}}
                           {1 - 2\xi\phi^2}\, d\phi \nonumber\\
             &=& \frac{1}{\sqrt{2}} \sqrt{\xi^{-1} - \xi_c^{-1}}\ \text{arcsin}
	           \left[ \sqrt{2} \sqrt{\xi - \xi_c^{-1} \xi^2}\, \phi \right]
	         + \frac{1}{\sqrt{2 \xi_c}}\ \text{arcsinh} \left[
		   \frac{\sqrt{2 \xi_c^{-1}}\, \xi\phi}{\sqrt{1 - 2\xi\phi^2}}
		   \right],
\end{eqnarray}
where
\begin{equation}
  \xi_c = \frac{1}{4}\, \frac{n-2}{n-1}.
\end{equation}
Particularly interesting is the case of conformal coupling $\xi=\xi_c$
because field redefinition
\begin{equation}
  \hat{\phi} = \frac{1}{\sqrt{2\xi_c}}\ \text{arctanh}
                 \left[ \sqrt{2\xi_c}\, \phi \right]
\end{equation}
can be inverted explicitly to give the recipe for obtaining conformally
coupled solutions from minimally coupled ones. It is
\begin{equation}
  \phi = \frac{1}{\sqrt{2\xi_c}}\, \tanh 
            \left[ \sqrt{2\xi_c}\, \hat{\phi} \right],
\end{equation}
\begin{equation}
  g_{\mu\nu} = \frac{\hat{g}_{\mu\nu}}{(1 - 2\xi_c\phi^2)^{2/(n-2)}}
             = \cosh^{4/(n-2)} \left[ \sqrt{2\xi_c}\, \hat{\phi} \right]
	         \hat{g}_{\mu\nu}.
\end{equation}
In particular, four dimensional Roberts solution becomes
\begin{equation}
  \phi = \sqrt{3} \tanh \left[ \frac{1}{\sqrt{3}}\,
         \text{arctanh}\, \frac{2pz-1}{\sqrt{1+4p}} \right],
\end{equation}
\begin{equation}
  ds^2 = \cosh^2 \left[ \frac{1}{\sqrt{3}}\,
         \text{arctanh}\, \frac{2pz-1}{\sqrt{1+4p}} \right]
	 \left\{-2 du\,dv + (u^2-uv-pv^2)\, d\omega^2\right\}.
\end{equation}
in conformally coupled model. This last expression was considered in
\cite{Oliveira&Cheb-Terrab:96}.

\subsubsection{Dilaton gravity} \label{sec:coupling:dilaton}

Another useful example is dilaton gravity described by the action
\begin{equation}
  S = \frac{1}{16\pi} \int \sqrt{-g}\, d^n\!x\, e^{-2\phi} \left[
        R + 4 (\nabla\phi)^2
      \right].
\end{equation}
Substituting coupling factors $F=e^{-2\phi}$, $G=-4 e^{-2\phi}$ into
relationship (\ref{eq:coupling:redefinition}), one can see that the
scalar field redefinition is a simple scaling
\begin{equation}
  \hat{\phi} = \sqrt{\frac{2}{n-2}}\, \phi, \hspace{1cm}
  \phi = \sqrt{\frac{n-2}{2}}\, \hat{\phi},
\end{equation}
and metrics differ by exponential factor only
\begin{equation}
  g_{\mu\nu} = \exp \left[ \sqrt{\frac{2}{n-2}}\, 2\hat{\phi} \right]
                 \hat{g}_{\mu\nu}.
\end{equation}
In particular, four dimensional Roberts solution becomes
\begin{equation}
  \phi = \text{arctanh}\, \frac{2pz-1}{\sqrt{1+4p}},
\end{equation}
\begin{equation}
  ds^2 = e^{2\phi} \left\{-2 du\,dv + (u^2-uv-pv^2)\, d\omega^2\right\}.
\end{equation}
in dilaton gravity.

\section{Conclusion} \label{sec:conclusion}

We have searched for and found continuously self-similar spherically
symmetric solutions of minimally coupled scalar field collapse in $n$
dimensional spacetime. For spacetime dimensions higher than three they
form one-parameter family and display critical behavior much like the
Roberts solution. Qualitative picture of field evolution is easy to
visualize in analogue with particle traveling in potential of
upside-down U shape. Unfortunately, the solutions in dimensions higher
than four can only be obtained in implicit form. Critical solutions are
in general simpler than other members of the family due to the
potential factoring. Strong similarity between Roberts solution and its
higher dimensional generalizations allows one to conjecture that these
higher dimensional critical solutions are not attractors either.
Absence of non-trivial self-similar solution in three dimensions raises
question whether scalar field collapsing in three dimensions exhibits
critical behavior at all. Perhaps further numerical simulations will
answer it.

We also use equivalence of scalar field couplings to generalize
solutions of minimally coupled scalar field to much wider class of
couplings. For often-used cases of conformal coupling and dilaton
gravity the results are remarkably simple. Some results of
\cite{Hirschmann&Wang:98}, applied for single scalar field only, become
trivial in view of this coupling equivalence.

However, the question of critical behavior of these generalized
solutions is complicated by the fact that the conformal factor $\Omega$
relating metrics (\ref{eq:coupling:metric}) for minimally and generally
coupled solutions may be singular. In the simplest case of non-singular
conformal transformation (i.e.\ when coupling $F$ is bounded and lower
bound is greater than zero) global properties of the minimally coupled
solution are preserved, and all important features of near-critical
collapse carry over on the generalized solution unchanged. If conformal
transformation is suspected to be singular, more careful study of global
properties of generalized solution (\ref{eq:coupling:redefinition}) is
necessary.

\section*{Acknowledgments}

This research was supported by Natural Sciences and Engineering
Research Council of Canada and Killam Trust.




\end{document}